\documentclass[aip,preprint]{revtex4-1}

\draft 

\usepackage{float}
\usepackage{siunitx}
\usepackage{upgreek}
\usepackage{graphicx}
\usepackage{caption}
\usepackage{subcaption}
\usepackage{ragged2e} 
\DeclareCaptionJustification{justified}{\justifying}

\usepackage[dvipsnames]{xcolor}

\usepackage{url,hyperref, lineno,microtype,subcaption}
\hypersetup{
    colorlinks=true,
    linkcolor=black,
    citecolor=blue,
    filecolor=black,      
    urlcolor=blue
}

\begin{document}


\title{
Dislocation cartography: Representations and unsupervised classification of dislocation networks with unique fingerprints}



\author{Benjamin Udofia}
\email{benjamin.udofia@rub.de}
\affiliation{Interdisciplinary Centre for Advanced Materials Simulation (ICAMS), Materials Informatics and Data Science, Ruhr-Universit\"at Bochum, Universit\"atsstr. 150, 44801 Bochum, Germany}

\author{Tushar Jogi}
\email{t.jogi@mpie.de}
\altaffiliation[Currently at ]{Max-Planck Institute for Sustainable Materials, Max-Planck Strasse 1, 40237 D\"usseldorf, Germany}

\author{Markus Stricker}
\email{markus.stricker@rub.de}
\affiliation{Interdisciplinary Centre for Advanced Materials Simulation (ICAMS), Materials Informatics and Data Science, Ruhr-Universit\"at Bochum, Universit\"atsstr. 150, 44801 Bochum, Germany}


\date{\today}


\begin{abstract}
Detecting structure in data is the first step to arrive at meaningful representations for systems.
This is particularly challenging for dislocation networks evolving as a consequence of plastic deformation of crystalline systems.
Our study employs Isomap, a manifold learning technique, to unveil the intrinsic structure of high-dimensional density field data of dislocation structures from different compression axis.
The resulting maps provide a systematic framework for quantitatively comparing dislocation structures, offering unique fingerprints based on density fields.
Our novel, unbiased approach contributes to the quantitative classification of dislocation structures which can be systematically extended.
\end{abstract}


\maketitle 


\section{Introduction}

Since the first proposal of dislocations as the carrier of plastic deformation\cite{Taylor1934,Polanyi1934,Orowan1934a} and subsequent experimental confirmation,\cite{Menter1956, Hirsch1956} many efforts went into characterizing dislocation structures under different conditions to better understand the relationship between loading, motion, dislocation structure, and macroscopic properties:
Cyclic loading leads to dislocation patterning with wall and cell structures \cite{Mughrabi1983,Hussein2016} or persistent slip bands \cite{Kubin2016} as well as certain local configurations of dislocations.\cite{El-Achkar2018}
Indentation loading results in specific patterns, depending on the indenter geometry and surface orientation, \cite{Gagel2016} as does bending \cite{Motz2012}, torsion \cite{Ziemann2015,Stricker2022} or tension loading in different crystallographic orientations \cite{Stricker2015a} which favor or oppose different configurations, depending on activation of slip systems and their interaction through junctions. \cite{Sills2018,Stricker2018,Akhondzadeh2021}
Body-centered cubic crystals show different structures because of the difference in edge and screw dislocation character mobility; \cite{Weygand2015} microstructures with multiple phases like those of superalloys result in yet another dislocation structure. \cite{Gao2015,jogi2021interfacial}
In addition, dislocation mobility and, therefore, structure is affected by temperature \cite{Chen2016} or local chemical order \cite{Sudmanns2021} as well as the aspect ratio in small scale specimen. \cite{Senger2011}

It is undisputed that the macroscopic properties are a function of the (local) dislocation structure(s). \cite{suzuki2013dislocation,Stricker2018,Akhondzadeh2023,Deka2023}
Consequently, substantial efforts have gone into both experimental \cite{Zhang2022,Sills2022,Yildirim2023,Wang2023a} and computational \cite{Steinberger2016,Steinberger2016a,Demirci2023,Katzer2024} characterization and to develop strategies for their representation. \cite{Robertson2021}
In its essence, this is the search for variables rather than confounding variables.\cite{Hiemer2023}
Discrete dislocation dynamics simulations provide a wealth of complex, evolving network data, however it is notoriously difficult to analyze despite intense mathematical efforts with a hierarchy of tensorial quantities to capture geometry and topology of dislocation networks.\cite{Weger2021}
Notably, Steinberger et al.~\cite{Steinberger2016a} developed a scheme which establishes a correlation between subvolumes of discrete dislocation simulations based on such tensorial features derived from continuum dislocation dynamics theory (CDD) \cite{Hochrainer2014} and specimen size.
More specifically, they chose classes based on specimen sizes of simulations from which the (sub)volume was originally extracted.
This approach, while insightful for predicting specimen sizes, may not be as directly applicable when characterizing the complex features of specific dislocation structures themselves.
Another approach for the representation of dislocations is the development of a dislocation ontology, called DISO. \cite{ihsan2023diso}
However, this also does not solve the problem of characterizing dislocation structures systematically.
Despite all efforts, the quantitative assessment of similarities and differences between dislocation structures from different loading scenarios, orientation, crystal structures, etc., remains a challenge and requires a methodological advancement.

To address this challenge of quantitative dislocation structure characterization, we present a methodology for uniquely creating fingerprints of individual dislocation structures.
These fingerprints not only serve as unique identifiers but also encode the intricate relationships between each other.
An analogy would be atomistic systems.
For these, specifically in the context of the development of machine-learned interatomic potentials based on appropriate representations, this has largely been solved~\cite{Deringer2021,Musil2021b,Qamar2023} and much of the effort of the community is focused on efficient implementations.~\cite{Bochkarev2022}
No such convergence of agreement for representations exists for dislocation systems.
Our contribution represents an initial step towards the formalization of the development of representations for dislocation structures.
In this first step, we propose an unbiased approach of characterizing dislocation density fields using the concept of embeddings.
Our method proves to be a powerful tool for the quantitative comparison of dislocation structures, enabling the identification of both differences and similarities.
Our approach is general and can be extended systematically to include other features of the dislocation network in addition or instead of the total dislocation density field we chose here.
We demonstrate the generality of our approach using two different discrete dislocation dynamics (DDD) codes and foresee that our approach accommodates data not only from simulations but potentially also from experimental sources, aligning with recent advances in experimental 3D characterizations of dislocations. \cite{Yildirim2023} Finally, our dislocation maps will be useful in machine learning approaches where a balanced training data set of diverse structures is crucial: our maps can be combined with unbiased sample selection strategies like farthest point sampling. \cite{Cersonsky2021}

The paper is organized with details about the generation of the data using discrete dislocation dynamics and the analysis using embeddings in section \ref{sec:methods}.
We then present the resulting dislocation structure maps in section~\ref{sec:result}, discuss implications of discretization, open vs. periodic boundary conditions, variations of compression axes, the choice of hyperparameters, and the concept of `similarity' of dislocation structures in section~\ref{sec:discussion}. 
The paper closes with concluding remarks about the implications and a future path towards the development of unbiased dislocation structure characterization in section~\ref{sec:conclusion}.

\section{Methods}
\label{sec:methods}
The basis for the dislocation network characterization is a dataset of discrete dislocation dynamics simulations, openly available on Zenodo including the script for interacting with it.\cite{udofia2024-11354118, Jogi2024zenodo}
For these, we will briefly introduce the methodical background and postprocessing steps: DDD output of the dislocation structure is a set of points and segments.
This is a consequence of discretizing curved lines in space and is unsuitable as input to manifold learning techniques because the number of nodes and segments varies across timesteps and simulations.
A solution to this is to convert curved dislocation lines into density fields with a user-specified discretization.
Subsequently, we introduce our approach for quantitative characterization employing the concept of manifold learning resulting in embeddings.

\subsection{Discrete dislocation dynamics}
DDD is a computational method for modeling the collective behavior of dislocations in materials including their motion and interactions.
In DDD, dislocations are represented as lines which are discretized using nodes and segments.
Additional inputs constitutes all possible interactions between dislocations such as junctions.
Here, we use two different implementations of DDD to create the dislocation structure dataset.
Specifically, ParaDiS \cite{arsenlis2007enabling} and the ``Karlsruhe Code". \cite{weygand2002aspects, weygand2005study}
The distinction between the two implementations is mainly w.r.t. surfaces: periodic vs. open surfaces.
In the following, we provide the parameters common to both frameworks and parameters which are specific to each implementation.

\subsection{Simulation setup}
\label{sec:SDS}

\begin{figure}
     \centering
     \begin{subfigure}[h]{0.4\textwidth}
         \centering
         \includegraphics[width=0.85\textwidth]{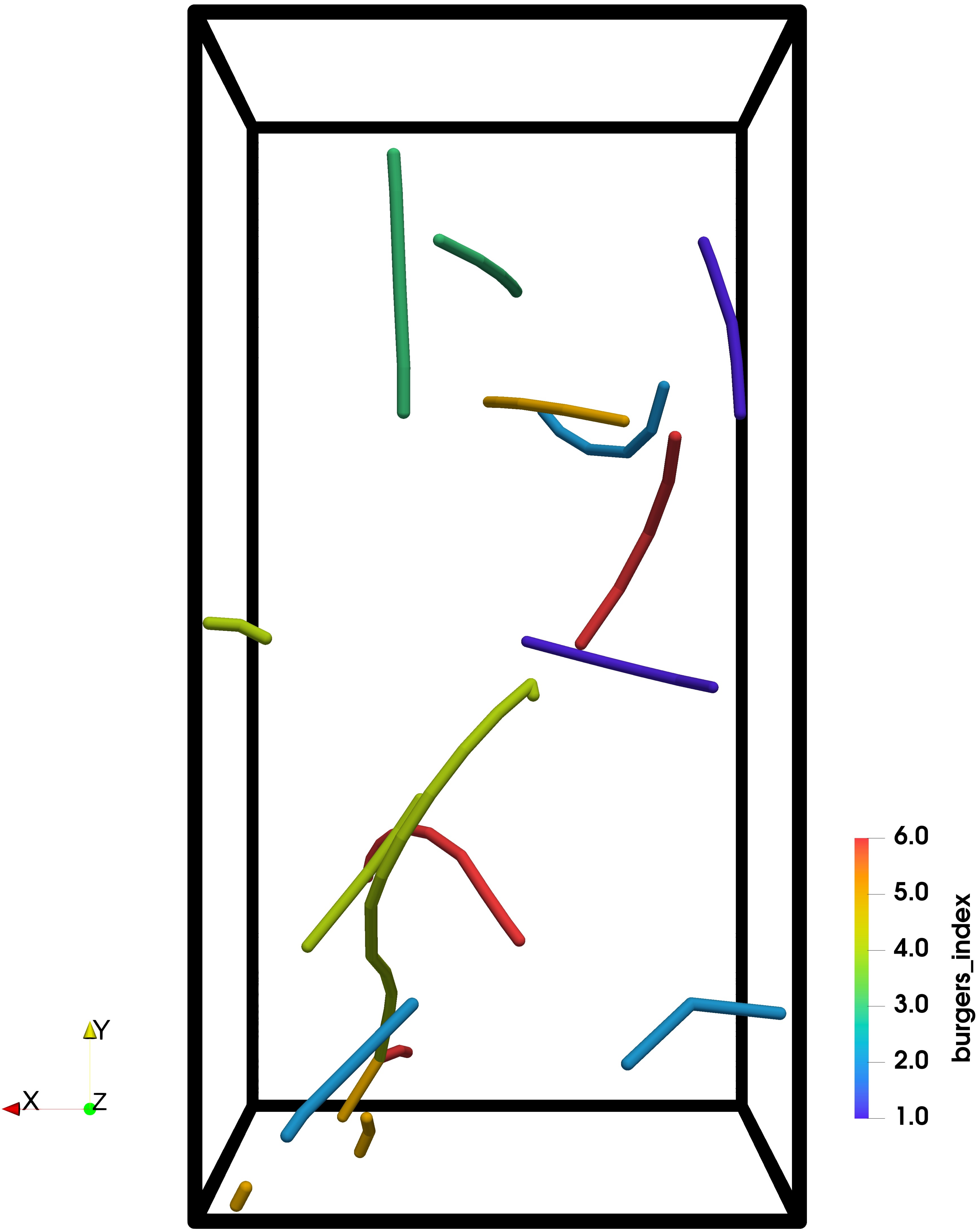}
         \caption{Karlsruhe Code (free surfaces).}
         \label{fig:OS}
     \end{subfigure}
     \hfill
     \begin{subfigure}[h]{0.55\textwidth}
         \centering
         \includegraphics[width=\textwidth]{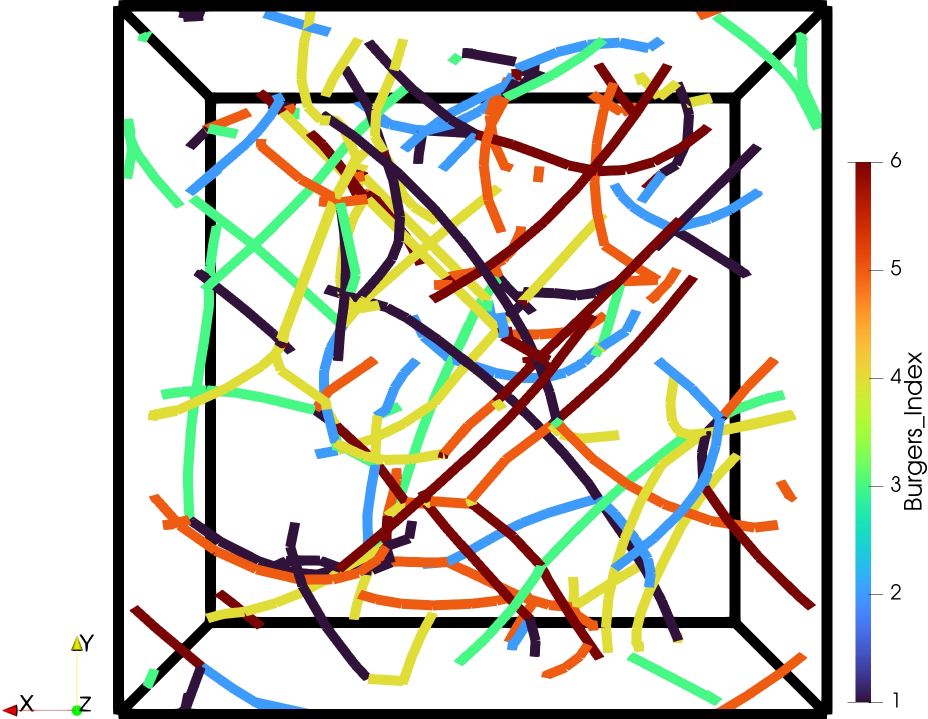}
         \caption{\label{fig:PBC} ParaDiS code (periodic boundary conditions).}
     \end{subfigure}
        \caption{\label{fig:1} Snapshot of simulated dislocation structures in perspective view initialized with randomly distributed Frank-Read sources. (a) Aspect ratio of $2$ for the open surface boundary condition with a low starting density of dislocation network. (b) Aspect ratio of $1$ for the periodic boundary condition with a high starting density of dislocation network. Strain rate-controlled displacement is applied in negative y direction.}
\end{figure}

\subsubsection{Common conditions}
All samples are compressed using displacement control with a constant strain rate of \SI{5000}{\per \second}, typical for DDD.
Compression loading was chosen based on the prevalent use of compression tests in small-scale plasticity experiments. \cite{frick2010plasticity, maass2012situ, weekes2015situ, malyar2017size}
Two input parameters are systematically varied: compression direction and initial dislocation density for a variety of resulting structures.
Compression is applied in directions $[100]$, $[110]$, $[111]$ and $[234]$ resulting in varying numbers of active slip systems and distinct dislocation structures.
Additionally, we explore deviations from perfect orientations and adjust the compression axis by increments of $\pm$\SI{5}{\degree} and $\pm$ \SI{10}{\degree}.
Density variations are a `low' initial density at \SI{2.5e13}{\per\meter\squared} and a `high' initial density of \SI{1e14}{\per\meter\squared}.
With this, we capture both the plastic regime characterized by the activation of a single source and the multiplication regime where dislocation interactions dominate the network evolution.
The resulting dislocation structures are then postprocessed into dislocation density fields with varying discretization.
Grid size of the density field is approximately the average dislocation distance $\bar{d} = \rho^{-1/2}$ at the end of the compression simulations.

\subsubsection{Material parameters for open surfaces}
For the open boundary condition, we use material parameters mimicking face-centered cubic (fcc) isotropic aluminum: a lattice parameter of $a =\SI{0.404}{\nano\meter}$, Poisson’s ratio $\nu = 0.347$, and shear modulus $G= \SI{27}{\giga\pascal}$.
The box size of the fcc base sample measures $240 \times 480 \times 240$ \si{\nano\metre}$^3$, an aspect ratio of $2$.
The initial dislocation structure consists of randomly distributed Frank–Read sources with a length of \SI{100}{\nano\meter}, subject to a variation of $\pm 20$\% within the volume (cf.~Fig.~\ref{fig:1}).
All simulations were compressed to a total strain of approximately $1$\%. 

\subsubsection{Material parameters for periodic boundary conditions}
For the periodic boundary condition (PBC), material parameters corresponding to pure copper are used: $a =$ \SI{0.361}{\nano\meter}, $G =$ \SI{56.9}{\giga\pascal}, and $\nu = 0.324$. Frank–Read sources were distributed within a cube-shaped volume with an aspect ratio of one (see Fig.~\ref{fig:1}). 
Simulations were conducted using a box size of $500 \times 500 \times 500$ \si{\nano\meter}$^3$, containing Frank-Read sources with lengths ranging from \SIrange{256}{281}{\nano\meter}. All simulation are performed up to total compressive strains ranging from \SIrange{0.2}{4}{\%} for different loading directions. 

\subsubsection{Postprocessing: density fields}
All the simulations were postprocessed into density fields to achieve a fixed input size for the manifold learning algorithms and facilitate meaningful comparisons among the different simulations.
For simulations conducted under periodic boundary conditions, discretizations of $10 \times 10 \times 10$ and $20 \times 20 \times 20$ were applied to the density fields, resulting in resolutions of \SI{50}{\nano \meter} and \SI{25}{\nano \meter}, respectively.
For open boundary simulations we used discretizations of $8 \times 16 \times 8$ and $16 \times 32 \times 16$, resulting in spatial resolutions of \SI{30}{\nano \meter} and \SI{15}{\nano \meter}, respectively.

\subsection{Embeddings through manifold learning}
The field representations of the previous step of dislocation structures result in a high-dimensional representation ranging from $1000$ to $8192$ values depending on discretization.
Our objective is to find \textit{structure} in these datasets for effective characterization and visualization of the similarities and differences resulting from different initializations, compression axes, and open vs. periodic surfaces.
We therefore apply several manifold learning techniques to our dataset to create mappings from the high-dimensional dislocation density fields to the low-dimensional embeddings, here called dislocation structure maps.
These embeddings serve as a representation of a specific dislocation structure.
Manifold learning techniques have become a cornerstone of modern machine learning, offering a powerful approach to represent data in a more informative and compact form.
In a general sense, embeddings are a mathematical representation of an object or entity in a lower-dimensional space while preserving the relationship to other entities in the dataset.
The goal is to capture the important features, characteristics, or relationships of objects in a condensed form that can be easily processed by algorithms.
The basic idea behind embeddings is to reduce complex and usually high-dimensional representations of objects in a simplified and compact form, which can be used for efficient computation and analysis, such as data visualization, \cite{iwata2004parametric} dimensionality reduction, \cite{roweis2000nonlinear} and machine learning. \cite{lecun2015deep}
Embeddings are used in a wide range of fields, including computer vision, \cite{sitzmann2019deepvoxels} graph analysis, \cite{goyal2018graph} and recommendation systems. \cite{zhang2016collaborative}

We employ the manifold learning technique Isometric Mapping for the generation of two-dimensional embeddings of the high-dimensional dislocation density dataset.
We choose this algorithm based on superior performance and physical interpretability compared to other embedding schemes (cf. Supplementary Fig. 1 for comparison with other algorithms we assessed).

\subsubsection*{Isomap embeddings}
Isomap is a non-linear dimensionality reduction method designed to capture the intrinsic geometry of high-dimensional data.
It constructs a weighted graph based on nearest neighbors with edges representing distances.
Geodesic distances between all pairs of points are estimated on the graph, and classical multidimensional scaling is employed to embed the data into a lower-dimensional space, preserving the underlying structure of non-linear manifolds.
The choice of the number of nearest neighbors is a critical factor that significantly influences the outcomes, exerting a substantial impact on the effectiveness of Isomap.
This approach has been widely used in various fields to reveal the inherent geometry and relationship of complex entities, offering insights into their underlying structure and relationship.\cite{tenenbaum2000global}

Isomap can be conceptualized as an extension of multi-dimensional scaling (MDS) or kernel principal component analysis (PCA). \cite{kruskal1964nonmetric, borg2005modern}
The Isomap algorithm, for instance, implemented in the popular library \texttt{scikit-learn}, \cite{scikit-learn} navigates through three distinct stages: locating nearest neighbors, uncovering shortest paths in graphs, and executing partial eigenvalue decomposition. Computational complexity is characterized by input dimensions, neighbor count, and desired output dimensions. \cite{tenenbaum2000global}

\section{Results} \label{sec:result}


Embedding maps of the dislocation microstructures for the various loading directions are shown in Figures \ref{fig:LR-LD} to \ref{fig:PBC-HR}.
We present the embedding maps along two general tracks: open vs. periodic boundaries.
Within the open boundary section we compare different activations by compression along the $[100]$, $[110]$, $[111]$, and $[234]$ crystallographic directions of fcc crystals.
Along these orientations, we vary the initial dislocation density and the coarse vs. fine discretization of the density field.
In addition, we present embedding maps for compression simulations varying $\pm \SI{5}{\degree}$ and $\pm \SI{10}{\degree}$ from the perfect crystallographic alignment of the compression axes.
For the periodic boundary conditions, we present results for perfectly aligned compression axes as above using a coarse discretization for the low-density initialization and a fine discretization for the high-density initialization.
The periodic boundary conditions case serves to show that our analysis is generally applicable to dislocation structures.

\subsection{Open surfaces}


\subsubsection{Low resolution}
\begin{figure}[htp!]
     \centering
     \begin{subfigure}{0.45\textwidth}
         \centering
        \includegraphics[width=\linewidth]{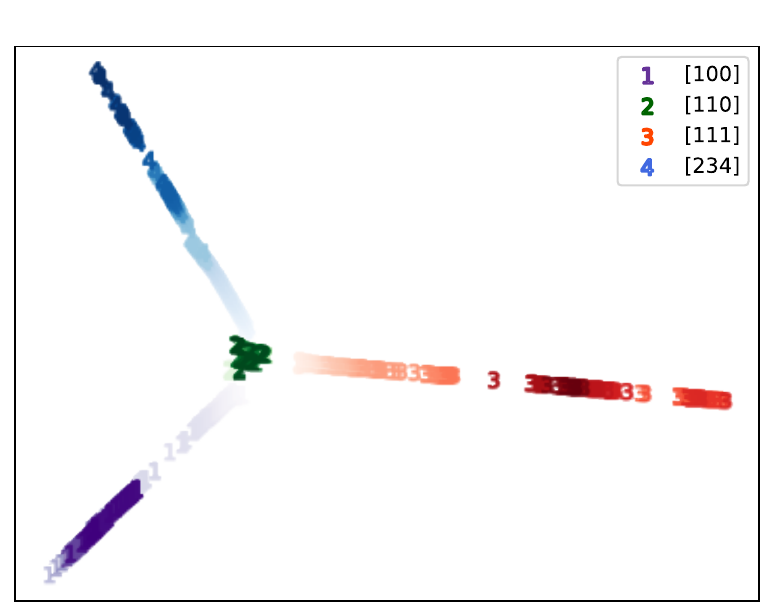}
         \caption{\label{fig:LR-LD} Low density case.}
     \end{subfigure}
     \hfill
     \begin{subfigure}{0.45\textwidth}
         \centering
        \includegraphics[width=\linewidth]{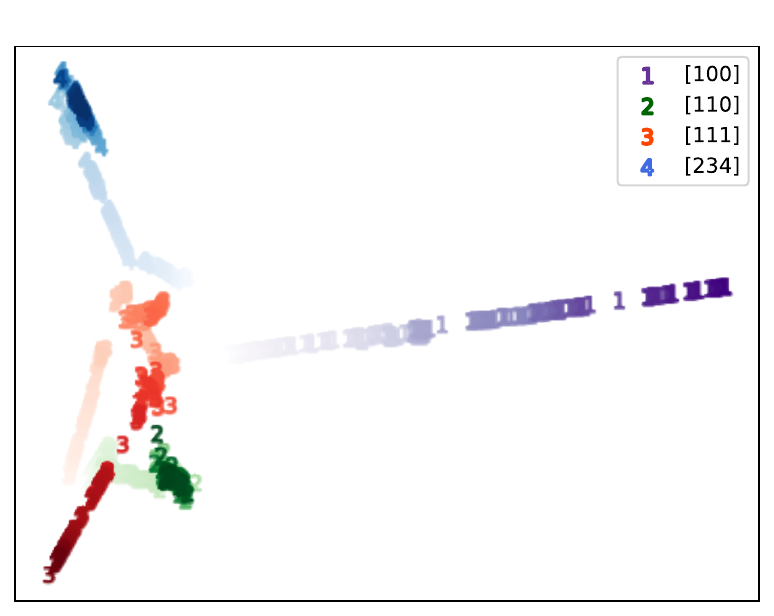}
         \caption{\label{fig:LR-HD} High density case.}
     \end{subfigure}
        \caption{\label{fig:LR} Isomap embedding maps for perfect crystal orientations at a discretization of $8 \times 16 \times 8$, with $k$ nearest neighbors set to $30$. (a) Initial density of \SI{2.5e13}{\per\meter\squared}. (b) Initial density of \SI{1e14}{\per\meter\squared}.}
\end{figure}

Figure~\ref{fig:LR} shows embeddings maps for simulations with the compression axes perfectly aligned with crystallographic directions using the coarse discretization density fields.
The low density map (Fig.~\ref{fig:LR-LD}) shows three diverging orientations with increasing total strain (faint to dark color) from a common central point indicating a more similar structure at initialization and dissimmilar structures after compression. 
The $[110]$ compression axis trajectory stays at the common central starting point.
The high density map (Fig.~\ref{fig:LR-HD}) shows a less pronounced common starting point but also a generally diverging behavior with increasing total strain.

\subsubsection{High resolution}
\begin{figure}[htp!]
     \centering
     \begin{subfigure}{0.46\textwidth}
         \centering
        \includegraphics[width=\linewidth]{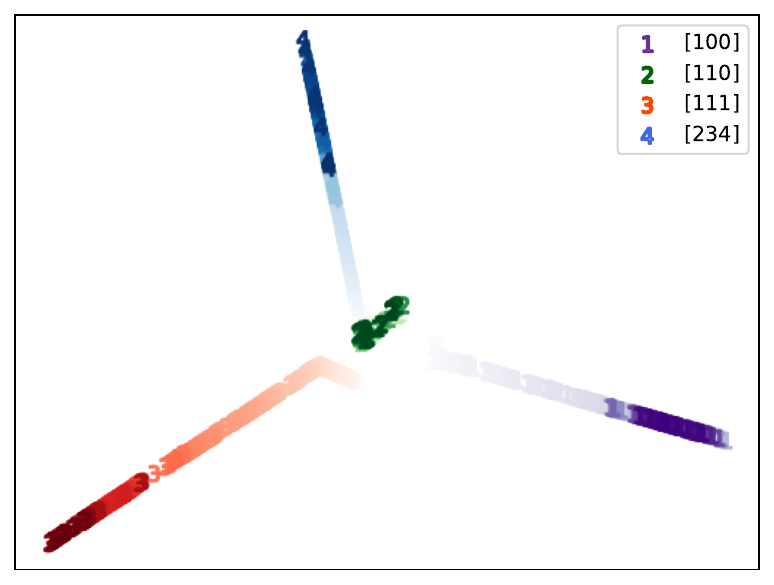}
         \caption{\label{fig:HR-LD} Low density case.}
     \end{subfigure}
     \hfill
     \begin{subfigure}{0.46\textwidth}
         \centering
        \includegraphics[width=\linewidth]{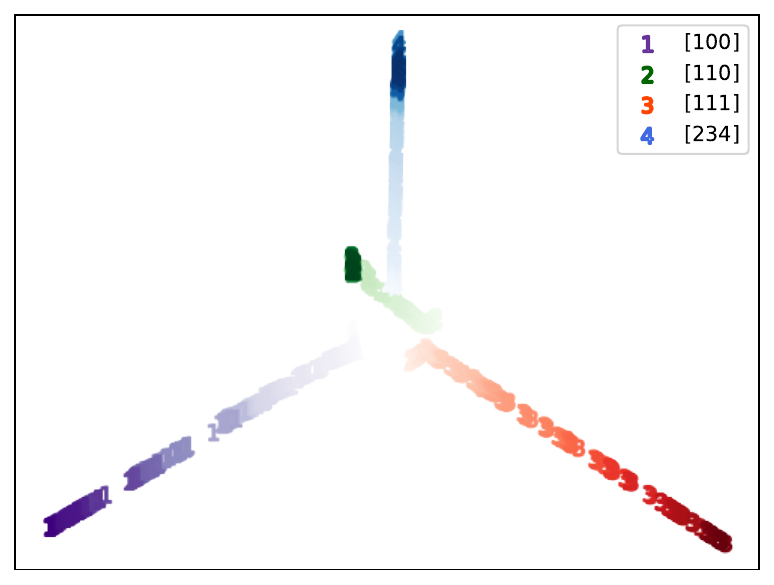}
         \caption{\label{fig:HR-HD} High density case.}
     \end{subfigure}
        \caption{\label{fig:HR} Isomap embedding results for perfect crystal orientations at a discretization of $16 \times 32 \times 16$, with $k$ nearest neighbors set to $50$. (a) Initial density of \SI{2.5e13}{\per\meter\squared}. (b) Initial density of \SI{1e14}{\per\meter\squared}.}
\end{figure}

Figure~\ref{fig:HR} shows the same data as Figure~\ref{fig:LR}, however using a higher resolution dislocation density field to represent discrete dislocation structures from the low and high density initializations.
The previously observed diverging behavior from a common starting point is more pronounced, in particular for the high-density case.

\subsubsection{Variations in crystal orientation}
\begin{figure}[ht]
     \centering
     \begin{subfigure}{0.46\textwidth}
         \centering
        \includegraphics[width=\linewidth]{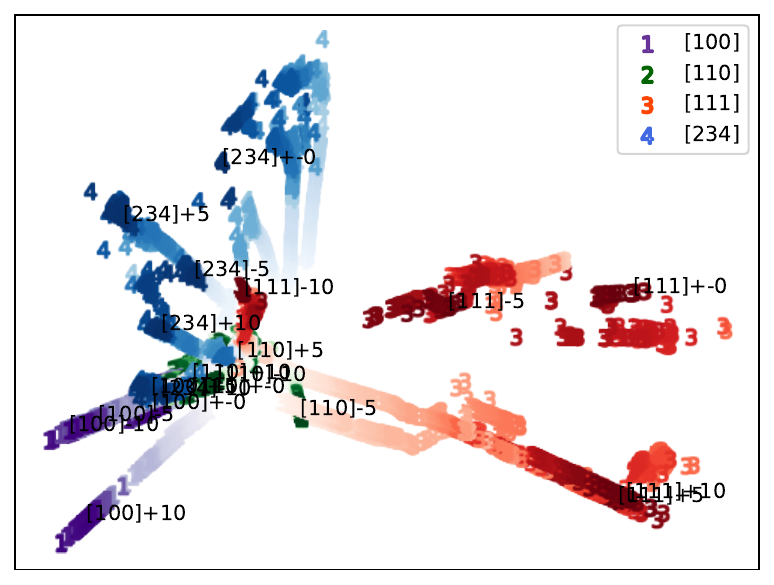}
         \caption{\label{fig:CM-LR-LD} Low density case.}
     \end{subfigure}
     \hfill
     \begin{subfigure}{0.46\textwidth}
         \centering
        \includegraphics[width=\linewidth]{Images/16x32x16/hr_combined_isomap_1e14_n50_normal_deg.pdf}
         \caption{\label{fig:CM-HR-HD} High density case.}
     \end{subfigure}
        \caption{\label{fig:CM} Isomap embedding results for perfect crystal orientations ($\pm0$) and their deviations at angles of $\pm 5$ and $\pm10$ degrees. (a) Discretization of $8 \times 16 \times 8$, with $k$ nearest neighbors set to $30$. (b) Discretization of $16 \times 32 \times 16$, with $k$ nearest neighbors set to $50$.}
\end{figure}

In addition to the previously presented perfectly-aligned compression axes trajectories, Figure~\ref{fig:CM} shows trajectories with $\pm \SI{5}{\degree}$ and $\pm \SI{10}{\degree}$ deviation of the compression axes from the perfect crystallographic orientation.
Deviations from the crystallographic axes are indicated using individual labels.
In the low-density initialization case (Fig.~\ref{fig:CM-LR-LD}) with coarse discretization fields, similarly-oriented compression axes cluster together but vary. E.g. the $[100]+10$ case is closer to the $[110]-5$ case than to the $[234]+10$ case (cf. also Supplementary Fig. 2). 
In addition, similarities within one orientation group $[111]+5$, $[111]+10$, etc. are visible on the lower right, indicating that similar dislocation structures develop despite the variation in compression axes.
The high-density case (Fig.~\ref{fig:CM-HR-HD}) shows three strongly diverging trajectories, $[111]+-0$, $[100]+-0$, and $[111]-10$, whereas the other trajectories cluster in the region around the common starting point.

\subsection{Periodic boundary conditions}
\begin{figure}[htp!]
     \centering
     \begin{subfigure}{0.46\textwidth}
         \centering
        \includegraphics[width=\linewidth]{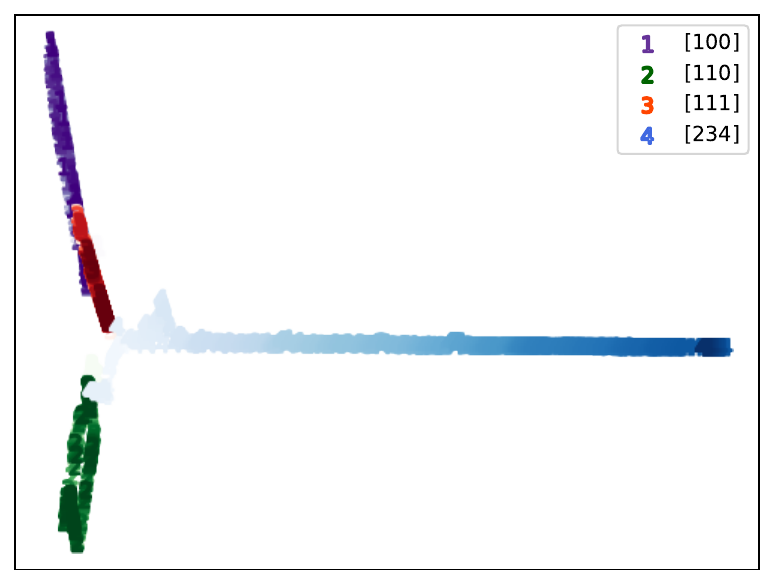}
         \caption{\label{fig:PBC-LR} Low density case.}
     \end{subfigure}
     \hfill
     \begin{subfigure}{0.46\textwidth}
         \centering
        \includegraphics[width=\linewidth]{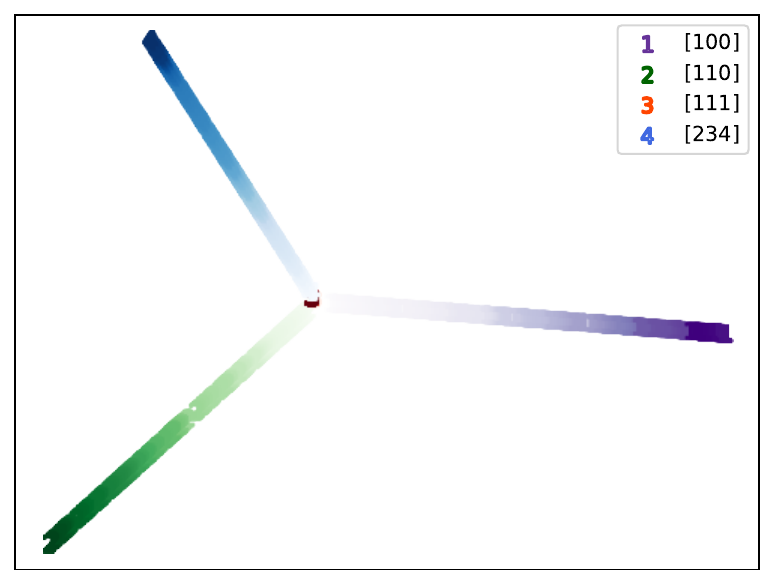}
         \caption{\label{fig:PBC-HR} High density case.}
     \end{subfigure}
        \caption{\label{fig:PBC-LR-HR} Isomap embedding results for perfect crystal orientations under PBC. (a) Discretization of $10 \times 10 \times 10$, with $k$ nearest neighbors set to 30. (b) Discretization of  $20 \times 20 \times 20$, with $k$ nearest neighbors set to 50.}
\end{figure}

In addition to the application of the manifold learning techniques to simulations using open surfaces, we also applied it to trajectories with periodic boundary conditions.
This was done to facilitate comparison with other simulation tools or frameworks.
The maps were generated for both the low-resolution, low-density case Fig.~\ref{fig:PBC-LR} and the high-resolution, high-density case Fig.~\ref{fig:PBC-HR}.
Dislocation trajectories are grouped into four different clusters corresponding to the loading directions (see legends), which start at a common central point.
The trajectories then distinctly deviate with increasing total strain (faint to darker colors).
In the low-density case (Fig.~\ref{fig:PBC-LR}), the $[111]$ and $[100]$, the directions overlap significantly; in the high-density case (Fig.~\ref{fig:PBC-HR}), all but the $[111]$ trajectory diverge whereas the latter stays close to the common starting point.

\section{Discussion}
\label{sec:discussion}

Embedding coordinates are a novel representation of dislocation networks and provide a unique view on the structural relationship of different discrete dislocation structures from various trajectories.
Clear correlations are visible within individual trajectories (cf. Fig~\ref{fig:LR}) as sequential simulation steps are mapped in sequence.
Further, different compression axes which lead to different glide system activations generally have different embedding coordinates in the maps.
Embeddings provide an unbiased numerical value for characterization and comparison of dislocation structures.
However, as with most techniques the outcome depends on several user-supplied choices.

It starts with the choice of representation of a discrete dislocation structure in its original form of nodes and segments from DDD.
A conversion step is necessary since any manifold learning technique requires the same number of dimensions for each datapoint to be used which is generally not the case for the representation of dislocation networks in form of segments and nodes.
In this first study, we chose the simplest form of field representation, dislocation density fields.
But any choice of representation of discrete dislocation structures in form of field variables, including expansions of higher-order alignment tensors,~\cite{Weger2021} should be explored in the future.
Note that for generality of the representation any currently existing choice of representation in form of continuum fields would still require some kind of rotation-invariance since this is generally not the case:~\cite{Suh2020}
Imagine exchanging the x and z axes in Figure~\ref{fig:1}, the dislocation structure would be the same, but its representation in form of a field would be markedly different.
A possible solution could be the use of correlations fields per structure.
In the following, we discuss our approach and its implications based on a fixed dataset as well as fixing the choice to represent dislocation structures as dislocation density fields.

\subsection{Discretization}
The results obtained from the low-resolution ($8 \times 16 \times 8$) discretization reveal a clear distinction in dislocation structures for the low-density case as depicted in Fig.~\ref{fig:LR-LD}, whereas such clarity is not evident for the high-density case (see Fig.~\ref{fig:LR-HD}). Conversely, the high-resolution ($16 \times 32 \times 16$ discretization) results, illustrated in Fig.~\ref{fig:HR}, distinctly delineate the dislocation structures for both the low and high-density cases.
These findings highlight the significance of resolution in discerning the differences in dislocation structures. Specifically, the $8 \times 16 \times 8$ discretization proves adequate for distinguishing dislocation structures in the low-density case (see Fig.~\ref{fig:LR-LD}), while the $16 \times 32 \times 16$ discretization is necessary for the high-density case (see Fig.~\ref{fig:HR-HD}).
Furthermore, these conclusions align with the comparison of resolution sizes and the average dislocation distances.
By incorporating the maximum obtained dislocation density values of \SI{4.38e13}{\per\meter\squared} and \SI{1.93e14}{\per\meter\squared} into the average dislocation distance relation (see Section~\ref{sec:SDS}), we determined the average dislocation distances of \SI{0.15}{\micro \meter} and \SI{0.072}{\micro \meter} for simulations with low and high starting densities, respectively.
Notably, there is a fivefold difference in the low-density case between the voxel size of the low resolution (\SI{0.03}{\upmu m}) and the obtained average dislocation distance (\SI{0.15}{\upmu m}), as well as a similar relationship for the high-density case between the voxel size of the high resolution (\SI{0.015}{\upmu m}) and the obtained average dislocation distance (\SI{0.072}{\upmu m}).
The observed fivefold difference between voxel sizes and average dislocation distances provides a sufficient resolution to distinguish different dislocation structures.
A similar trend is observed for the PBC case, implying that the captured dislocation structures are accurately represented within the resolution limits.
Other field representations of the discrete representation, e.g. with higher dimensional tensors~\cite{Weger2021}, will likely work with lower resolutions.\cite{steinberger2019machine}

\subsection{Open surface vs. periodic boundary conditions}
Various embedding techniques are explored to reduce the dimensionality of the generated datasets.
Comparing results from other techniques (cf. Supplementary Fig. 1) demonstrate that the Isomap embedding approach is the most appropriate.
This is due to its ability to not only identify the distinct compression axes in the datasets but also classify the data points according to the simulation steps -- the Isomap result is amenable to physical interpretation.
The divergence of each crystal orientation (relative to the compression axis) from the simulation onset to the final simulation step indicates the progression of deformation across the sample, where the divergence is visually conveyed by a change in color intensity within each category.

The observed behavior in the low-density cases, depicted in both the low-resolution (Fig.~\ref{fig:LR-LD}) and high-resolution (Fig.~\ref{fig:HR-LD}) dislocation maps, can be attributed to the presence of a spiral Frank-Read source.
This source expands and exits the volume, only to re-emerge in a repetitive manner with increasing compression.
Consequently, self-similar structures emerge (see Supplementary Fig. 3 for a 3-dimensional embedding view), which are effectively captured.

In contrast, the high-density cases shown in Fig.~\ref{fig:LR-HD} (low resolution) and Fig.~\ref{fig:HR-HD} (high resolution) display a different behavior: an increasing dissimilarity over time with now repeating structures.
Deformation and evolution of the structure is dominated by  multiple dislocation interactions leading to the formation of intricate networks which are well captured in the embedding coordinates.

The maps obtained dislocation trajectories from a different DDD implementation with periodic boundary conditions (ParaDiS, see Fig.~\ref{fig:PBC-LR-HR}) results in similar trends for both low and high-density cases.
In Figure~\ref{fig:PBC-LR}, self-similar structures are observed over time in the low-density case, attributed to the cyclic motion of one or two activated Frank-Read sources within the volume, while Figure~\ref{fig:PBC-HR} illustrates distinct dislocation structures over time in the high-density case due to interactions many dislocations.
These findings are well aligned with results of the using open surface boundary conditions and demonstrate that the developed classification scheme is not dependent on the specific simulation method used and can be extrapolated to other data sources, such as experimental frameworks.

\subsection{Orientation variations}
We examined the effect of misorientations by creating dislocation maps for the perfectly oriented crystal with deviations of $\pm \SI{5}{\degree}$ and $\pm \SI{10}{\degree}$ from the crystallographically-aligned compression axis, for both low and high-density initializations.
In the low-density case with low resolution (Fig.~\ref{fig:CM-LR-LD}), microstructures of +5° and +10° misorientations from the $[111]$ orientation overlap, indicating similar dislocation structures.
Similarly, microstructures of $-\SI{10}{\degree}$ and $-\SI{5}{\degree}$ misorientations from the $[100]$ orientation overlap, suggesting similarity in their dislocation structures. The observed overlap is clearly visible in Supplementary Fig. 2.
This indicates that despite the different nomenclatures for misorientations of $+\SI{5}{\degree}$ and $+\SI{10}{\degree}$ in the $[111]$ orientation, and $-\SI{10}{\degree}$ and $-\SI{5}{\degree}$ in the $[100]$ orientation, the trajectories and evolving structures are similar and are correctly projected into embedding space as such.

The results for the high-density case with high resolution depict a complex dislocation map, where most dislocation structures are mapped to the center of the map (see Fig.~\ref{fig:CM-HR-HD}).
This suggests that the chosen resolution for the field representations is likely insufficient to distinguish dislocation structures resulting from different compression axes adequately.
It also underscores the need for additional descriptors such as curvature\cite{steinberger2019machine} or higher order alignment tensors\cite{Weger2021} to represent discrete dislocation networks as fields with fixed dimensions.

\subsection{Methods}
The dislocation maps obtained depend on the dataset, the embedding algorithm, and its hyperparameters.
They are unique to the current dataset, and even slight changes in any of these factors results in a different map.
The following subsections highlight the reasons behind the selections for the orientations and hyperparameters.
Additionally, we discuss constraints imposed on the embeddings.

\subsubsection{Compression axes}
\begin{figure*}[htp!]
\begin{center}
\includegraphics[width=\textwidth]{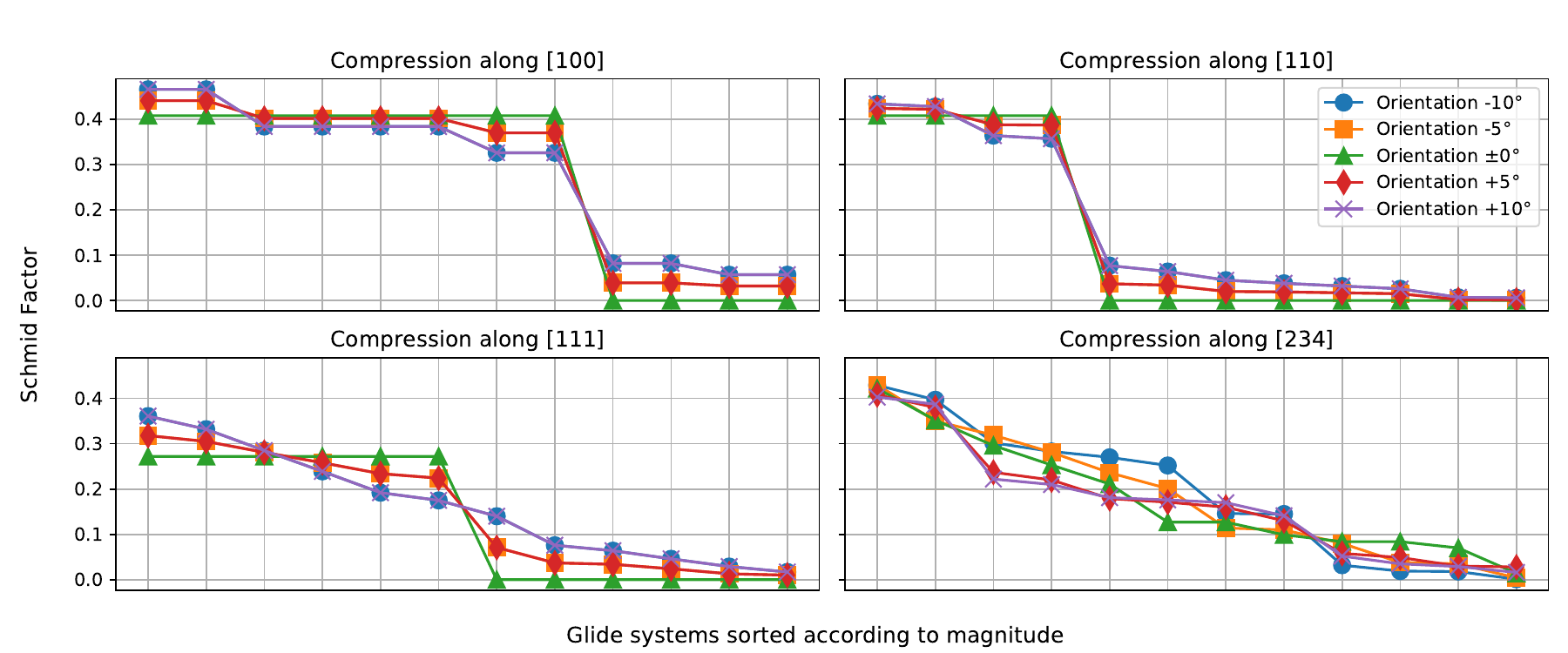}
\end{center}
\caption{Calculated Schmid factors for all 12 glide systems sorted according to magnitude: perfect orientations ($\pm\SI{0}{\degree}$) and  deviations of angles of $\pm\SI{10}{\degree}$ and $\pm\SI{5}{\degree}$.}\label{fig:schmid}
\end{figure*}

Figure~\ref{fig:schmid} shows the calculated Schmid factors for the analyzed perfect orientations (±0°) and their misorientations at angles of $\pm \SI{5}{\degree}$ and $\pm \SI{10}{\degree}$, sorted according to magnitude for the various slip systems.

Four crystal orientations were analyzed with respect to the compression axis.
These orientations encompass eight, six, four, and one dominant glide system(s) for the $[100]$, $[111]$, $[110]$, and $[234]$ orientations, respectively, as shown in Fig.~\ref{fig:schmid}, thereby ensuring the generation of markedly different microstructures. 

The Schmid factors order and magnitude illustrate the basic fact that a deviation from a given perfect orientation result in different active slip systems because the symmetry is disturbed. This observation is consistent with the findings shown in Fig.~\ref{fig:CM-LR-LD}, where the band widens in the embedding space as different misorientations are explored, suggesting that rotating away from the perfect orientation leads to a distinct distribution of the Isomap embeddings.

A \SI{-10}{\degree} variation in $[111]$ and $[234]$ compression as depicted in Figure~\ref{fig:schmid} shows a similar distribution, resulting in a comparable activation of dislocations.
This is further supported by the map illustrated in Figure~\ref{fig:CM-LR-LD}, where the dislocation structures for the -\SI{10}{\degree} variation in $[111]$ and $[234]$ compression have similar embedding coordinates:
We interpret the alignment of trajectories as a measure for the activation of similar glide systems, resulting in comparable dislocation structures.
However, this observation is also true for the +\SI{10}{\degree} variation in $[111]$ and $[234]$ compression, where a different distribution of the activated slip systems as shown in Fig. \ref{fig:schmid} results in very different dislocation microstructures (see Fig. \ref{fig:CM-LR-LD}). 
We attribute this similarity to our choice of field representation.
Therefore we expect that a field representation including more information about the dislocation structure in each voxel other than the total density would result in in a clearer separation in embedding space.

\subsubsection{Selection of hyperparameters}
The final choice of hyperparameters for the Isomap algorithm is based on an optimization to determine the ideal number of $k$-nearest neighbors.
We explored different values of the nearest neighbor parameter and observed that as the number of nearest neighbors increases, the dataset progressively separates into four distinct categories, representing various crystal orientations.
Beyond $k = 30$, these categories begin to exhibit similarities in the low-resolution case. 
In the high-resolution case, using the same parameters as for the low-resolution case ($k = 30$) results in maps that are less amenable to physical interpretation, as the $[110]$ orientation starts off from elsewhere and not the center on the map (see Supplementary Fig. 4).
Consequently, we determined the optimal number of nearest neighbors for achieving distinct categorization to be $k = 30$ and $k = 50$ for the low and high-resolution cases, respectively.
Optimization of the hyperparameters currently is a manual process.
In the future, this optimization should be automated and also based on data.

\subsubsection{Constraints}
One constraint imposed on the embeddings, informed by domain knowledge, is the expectation of similar dislocation structures at the starting point: each structure is initially populated with the same number of Frank-Read sources.
Conversely, it is anticipated that the end point will display markedly different dislocation structures across various orientations because they move on differently oriented crystallographic planes.

\begin{figure*}[htp!]
\begin{center}
\includegraphics[width=15cm]{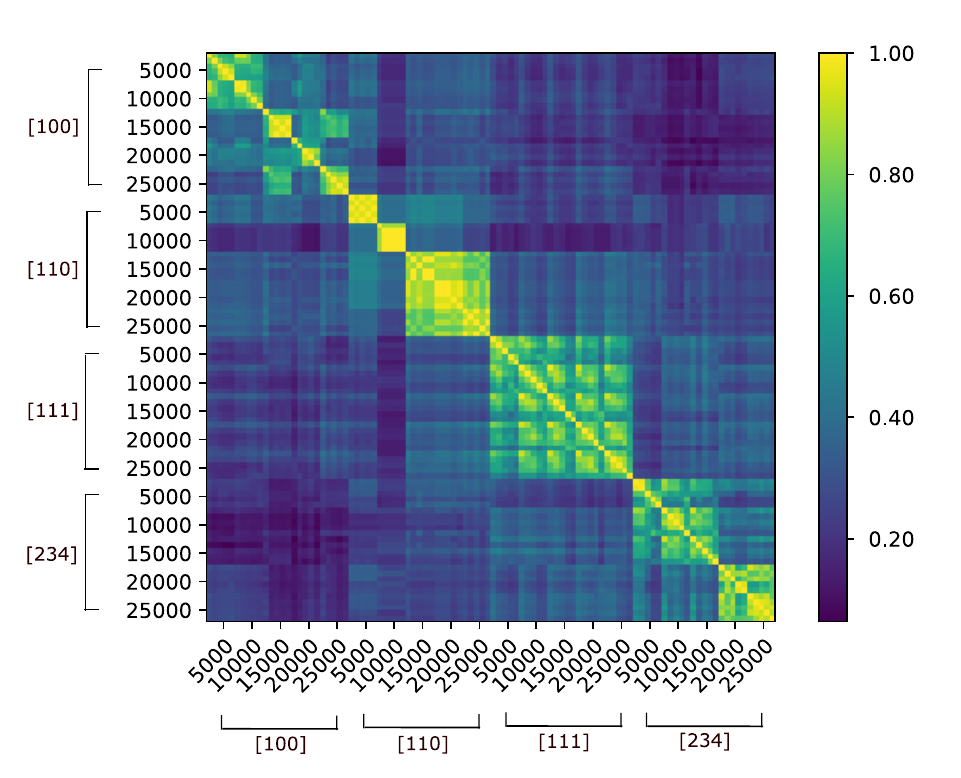}
\end{center}
\caption{\label{fig:cos-sim} Similarity matrix illustrating the cosine similarity between various orientations at simulation timesteps of 5000, 10000, 15000, 20000, and 25000 for the low-density low-resolution case.}
\end{figure*}

Figure~\ref{fig:cos-sim} shows a similarity matrix using the cosine similarity measure of the density fields for various crystal orientations ($[100]$, $[111]$, $[110]$, and $[234]$) at simulation steps 5000 , 10000, 15000, 20000, and 25000, corresponding to total strains of approximately 0.2 \%, 0.4 \%, 0.6 \%, 0.8 \%, and 1 \%, respectively.
Each simulation step comprises five pixels representing orientation variations: $\pm\SI{10}{\degree}$ $\pm\SI{5}{\degree}$ and $\SI{0}{\degree}$.
A comparison of the density fields of the perfect crystal orientations shows that within each orientation, the fields are more similar to each other than to those of the other orientations.
This observation aligns with the patterns in our embeddings in Section~\ref{sec:result}.
As a result, the manual selection of hyperparameters reflects our expectations regarding the distinctiveness of orientations in embedding space.

\section{Conclusion}
\label{sec:conclusion}

The main message of our study is that dislocation structures can be characterized and compared in novel ways:
Manifold learning techniques provide an unbiased and unsupervised view on inherent structure in data.
Finding and understanding inherent structure is the basis to establish representations which can be used to model complex systems.
This concept applies to dislocation systems as much as to any other system.
The most critical step before applying manifold learning techniques is the transformation of discrete dislocation systems represented as nodes and segments to a fixed dimension.
We choose the simplest form of a field representation in form of dislocation density fields.
However, \textit{ideal} representations are physically motivated, mathematically robust, exploit symmetries of the described system, efficient to calculate and uniquely describe the state of a system~\cite{Hiemer2023,Musil2021a}.
For atomistic systems with degrees of freedom of atom positions, chemistry, and unit cells in the context of interatomic potential development (force fields), this has largely been solved~\cite{Deringer2021,Musil2021b}.
Much of the effort of the community now revolves around their efficient implementation and training strategies.
For dislocation systems, no agreement exists on which continuum representation is the most appropriate for characterization and modeling.
In particular, because the complex networks of lines are notoriously difficult to analyze,~\cite{Weger2021} e.g. evidenced by the recent addition of evolution laws for junction point densities~\cite{Starkey2022} in the context of continuum dislocation dynamics~\cite{Hochrainer2014}.
Unbiased approaches like the one presented here can provide a new path to understanding the structure of complex dislocation networks which can then be correlated with their properties and used in modeling.

In our contribution we establish a methodology for quantitatively comparing dislocation structures, ultimately creating dislocation structure maps and, therefore necessarily, a ``coordinate system'' for dislocation structures.
Our approach lays the groundwork for building a \textit{navigation system} for the large variety of existing dislocation structures and their characterization.
The representative set of dislocation trajectories used here comprises several compression axes and densities as well as open and periodic boundary conditions.
Subsequently, the discrete structures were postprocessed into dislocation density fields which were then mapped using Isomap embeddings.

The findings of this study demonstrate the ability to differentiate evolving dislocation structures undergoing significant changes over time, particularly in the high-density case.
Additionally, the results highlight the capacity to quantify the similarity between dislocation structures that exhibit self-similarity despite evolving, as observed in the low-density case.
These resulting maps provide valuable insights into the dynamic nature of dislocation microstructures and their relationship and offer a means to characterize and compare their evolution.

Even with the simplest method of characterizing dislocation structure --- solely employing a scalar field of dislocation density on a limited dataset --- it is possible to extract maps that align with our (intuitive) physical understanding: distinct compression axes lead to distinct structures.
However, the sensitivity of the embeddings to (hyper)parameters implies that the representation of dislocation solely through density field descriptors should be improved.
Thus, it seems likely that more meaningful representations of dislocation networks including alignment tensors~\cite{Weger2021} or network properties based on graph measures into the representation of dislocation structures.
Our approach could ultimately provide a new path for the representation and modeling of dislocation systems.

\section*{Supplementary Material}
The supplementary material for this manuscript contains four figures. Supplementary Fig. 1 shows the results from all the manifold learning techniques explored in this research study. Supplementary Fig. 2 clearly depicts the distinct plots of each crystal orientation and loading direction, properly distinguishing the overlaps observed in Fig.~\ref{fig:CM-LR-LD}. Supplementary Fig. 3 presents a visualization of evolving self-similar structures in perfect crystal orientations using low-density, coarse discretization under open surface boundary conditions. Supplementary Fig. 4 presents the Isomap embedding result for the high-density, fine discretization case, using $k = 30$ as the number of nearest neighbors and open boundary conditions.

\begin{acknowledgments}
Benjamin Udofia and Markus Stricker gratefully acknowledge funding by the German Research Foundation (DFG) through project STR 1729/1-1, project number 469020538.
\end{acknowledgments}

\section*{Author Declarations}
\subsection*{Conflict of Interest}
The authors have no conflicts to disclose.
\subsection*{Author Contributions}

\textbf{Benjamin Udofia:} data curation (lead); formal analysis (lead),
methodology (equal); software (lead); validation (lead); visualization
(lead); writing -- original draft (lead); writing -- review and
editing (support).
\textbf{Tushar Jogi:} data curation (support); formal analysis (support),
methodology (support); software (support); validation (support);
visualization (support); writing -- original draft (support); writing
-- review and editing (support).
\textbf{Markus Stricker:} data curaction (support); formal analysis (support);
methodology (equal); project administration (lead); software
(support); supervision (lead); validation (support); visualization
(support); writing -- original draft (support); writing -- review and
editing (lead).

\section*{Data Availability Statement}
The data that support the findings of this study are openly available in Zenodo at \url{https://doi.org/10.5281/zenodo.11354118}, \cite{udofia2024-11354118} and \url{https://zenodo.org/doi/10.5281/zenodo.11154055}\cite{Jogi2024zenodo}. The development of the scripts for interacting with the data is accessible on GitLab: \url{https://gitlab.ruhr-uni-bochum.de/udofibzd/ddd_embeddings}.

\bibliography{ddd}

\end{document}